\documentclass[aoas,MSNbibl,nameyear,seceqn,dvips]{arximspdf}
\usepackage{dcolumn,multirow}
\usepackage{graphicx}

\doi{10.1214/12-AOAS539} %kopijuoti is PTS
\volume{6}
\issue{3}
\pubyear{2012}
\firstpage{1306}
\lastpage{1326}

\makeatletter
\newcolumntype{d}[1]{D{.}{.}{#1}}

\newtheorem{theorem}{Theorem}
\newproclaim{definition}{Definition}
\newproclaim{remark}{Remark}

\newcommand{\bE}{\mathbf{E}}
\newcommand{\bP}{\mathbf{P}}
\newcommand{\bY}{\mathbf{Y}}
\newcommand{\bmu}{\bolds{\mu}}
\newcommand{\btau}{\bolds{\tau}}
\newcommand{\argmin}{\operatorname{argmin}}

\makeatother

\begin{document}
\begin{frontmatter}

\title{The screening and ranking algorithm to detect DNA~copy number
variations\thanksref{M1}}
\runtitle{Screening and ranking algorithm}

\thankstext{M1}{Supported by the University of Arizona Internal grant
and National Institute on Drug Abuse Grant R01-DA016750.}

\begin{aug}
\author[A]{\fnms{Yue S.} \snm{Niu}\corref{}\ead[label=e1]{yueniu@math.arizona.edu}} %
\and
\author[B]{\fnms{Heping} \snm{Zhang}\ead[label=e2]{heping.zhang@yale.edu}}
%%\thanksref{m2}
\runauthor{Y. S. Niu and H. Zhang}
\affiliation{University of Arizona and Yale University}
\address[A]{Department of Mathematics \\
University of Arizona\\
Tucson, Arizona 85721\\
USA\\
\printead{e1}} %\printead*{e2}}

\address[B]{Department of Epidemiology\\
\quad and Public Health\\
Yale University\\
School of Medicine\\
New Haven, Connecticut 06520\\
USA\\
\printead{e2}}

\end{aug}

% HISTORY:
\received{\smonth{12} \syear{2010}}
\revised{\smonth{1} \syear{2012}}

% ABSTRACT
%
\begin{abstract}
DNA Copy number variation (CNV) has recently gained considerable
interest as a source of genetic variation that likely influences
phenotypic differences.
Many statistical and computational methods have been proposed and
applied to detect CNVs based on data that generated by genome analysis
platforms.
However, most algorithms are computationally intensive with complexity
at least $O(n^2)$, where $n$ is the number of probes in the experiments.
Moreover, the theoretical properties of those existing methods are not
well understood.
A faster and better characterized algorithm is desirable for the ultra
high throughput data. In this study, we propose the Screening and
Ranking algorithm (SaRa) which can detect CNVs fast and accurately with
complexity down to $O(n)$. In addition, we characterize theoretical
properties and present numerical analysis for our algorithm.
\end{abstract}

% KEYWORDS
%

\end{frontmatter}

%s1 ###
\section{Introduction}\label{sec1}

The problem of change-points detection has been studied by researchers
in various fields including statisticians, engineers, econo\-mists,
climatologists and biostatisticians since 1950s.
We refer to \citet{Bhattacharya1994} for an overview of this subject. In
this paper, we focus on a high dimensional sparse normal mean model
%e1 ###
%
\begin{equation}\label{model}
Y_i = \mu_i +\varepsilon_i, \qquad 1 \leq i \leq n,
\end{equation}
where $\bY=(Y_1,\ldots,Y_n)^T$ is a sequence of random variables, its mean
$\bmu=(\mu_1,\ldots,\mu_n)^T$ is a piecewise constant vector and the
errors $\varepsilon_i$ are i.i.d. $\sim N(0, \sigma^2)$. A change-point
in this model is a position $\tau$ such that $\mu_{\tau}\ne\mu
_{\tau
+1}$. Assume that there are $J$ change-points $0 <\tau_1< \cdots<
\tau
_J< n$. We are interested in the case when $n$ is large, and $J$ is
small. The goal is to estimate the number of change-points, $J,$ and
the location vector\vadjust{\goodbreak} $\btau=(\tau_1,\ldots,\tau_J)^T$. Note that all
positions are potential candidates for a change-point, which makes the
problem high dimensional and hard to solve. However, the small~$J$
implies the sparse structure of the model, which we can utilize to
develop fast and efficient algorithms.

%s1.1 ###
\subsection{Background}\label{sec1.1}
The main motivation of this project is to develop efficient statistical
techniques for DNA copy number variation (CNV) detection, for which
model~(\ref{model}) plays an important role, for example, \citet
{OVLW:04}, \citet{HWLZ:05}, \citet{ZhangSiegmund2007}, \citet{Tibshirani:Wang:2008}, among others.
The DNA copy number of a region is the number of copies of the genomic
DNA. CNV usually refers to deletion or duplication of a region of DNA
sequences compared to a reference genome assembly. Recent studies in
genetics have shown that CNVs account for a substantial amount of
genetic variation and provide new insights in disease association
studies [\citet{FreemanEtal:2006}, \citet{McCarroll:Altshuler:2007}].
Identification of CNVs is essential in the systematic studies of
understanding the role of CNV in human diseases. Over the last decade,
efforts have been made to detect DNA copy number variation with the
help of significant advances in DNA array technology, including array
comparative genomic hybridization (aCGH), single nucleotide
polymorphism (SNP) genotyping platforms, and next-generation
sequencing. We refer to \citet{ZhangEtal:CNV:2009} for a nice review of
these technologies. These various platforms and techniques have
provided a large amount of data that are rich in structure, motivating
new statistical methods for their analysis.

%s1.2 ###
\subsection{Current segmentation methods}\label{sec1.2}
Historically, the case that there is at most one change-point in model
(\ref{model}) has been intensively studied. See \citet
{SenSri}, \citet{JamesSiegmund}, among others.
The problem of single change-point detection is equivalent to the
following hypothesis testing problem:
%e2 ###
%
\begin{eqnarray}\label{onetest}
H_0 \dvtx \mu_1&=&\cdots=\mu_n,\qquad \mbox{against},
\nonumber
\\[-8pt]
\\[-8pt]
\nonumber
H_1 \dvtx \mu_1&=&\cdots=\mu_j\neq\mu_{j+1}=\cdots=\mu_n \qquad\mbox{for some }j.
\end{eqnarray}
Let $S_k$ be the partial sum of observations $\sum_{i=1}^k Y_i$. When
$\operatorname{Var}(Y_i)$ is known, a~commonly used test statistic is the log
likelihood ratio statistic
%e3 ###
%
\begin{equation}\label{SS}
T_n=\max_{1\leq j\leq n-1} \{(jS_n/n-S_j)^2/[j(1-j/n)]\}.
\end{equation}
The log likelihood ratio test is usually satisfactory although the
distribution of $T_n$ is quite complicated. We refer to \citet
{CsorgoHorvath:1997} for the theoretical analysis of the log likelihood
ratio test.

When there are multiple change-points, the problem is much more
complicated. Here we briefly review several popular approaches,
including an exhaustive search with the Schwarz criterion\vadjust{\goodbreak} [\citet
{Yao1988}, \citet{Yao89}], the circular binary segmentation method [\citet
{OVLW:04}, \citet{VO:07}] and the $\ell_1$ penalization method [\citet
{HWLZ:05}, \citet{Tib:fusedlasso:2005}, \citet{Tibshirani:Wang:2008}].

To estimate $J$ and $\btau$, an exhaustive search among all
possibilities $0\leq\tilde{J}\leq n-1$ and $0<\tilde\tau_1<\cdots
<\tilde\tau_{\tilde{J}}<n$ can be applied, theoretically. Let
$\tilde
{\btau}=(\tilde\tau_1,\ldots,\tilde\tau_{\tilde{J}})^T$. For each
candidate $(\tilde{J},\tilde{\btau})$, the MLE for the variance,
$\hat
\sigma^2_{(\tilde{J},\tilde{\btau})}$, can be calculated by the least
squares method. \citet{Yao1988}, \citet{Yao89} applied the Schwarz
criterion, or
the Bayesian Information Criterion (BIC), to estimate the number and
locations of change-points and showed the consistency of the estimate.
In particular, the number of change-points is estimated as
%e4 ###
%
\begin{equation}\label{JHat}
\hat{J}=\mathop{\argmin}_{\tilde{J}} \frac{n}{2}\log\hat{\sigma
}^2_{\tilde
{J}}+\tilde{J}\log n \qquad\mbox{where } \hat{\sigma}^2_{\tilde
{J}}=\min_{\tilde{\btau}}\hat\sigma^2_{(\tilde{J},\tilde{\btau})}.\vspace*{-2pt}
\end{equation}
Once $\hat{J}$ is determined, the location estimator is defined as
\[
\hat{\btau}=\mathop{\argmin}_{\tilde{\btau}}\hat\sigma^2_{(\hat
{J},\tilde{\btau})}.\vspace*{-2pt}
\]
Unfortunately, when $n$ is large, these estimators are not attainable
due to the computational complexity. \citet{BBM00} employed a dynamic
programming algorithm to reduce the computational burden to the order
of $O(n^2)$. But it is still computationally expensive for large $n$.

An alternative approach is through the binary segmentation method,
which applies the single change-point test recursively to determine all
the change-points. As pointed out in \citet{OVLW:04}, one of the
drawbacks is that the binary segmentation can hardly detect small
segments buried in larger ones. To overcome this problem, \citet
{OVLW:04} introduced a~Circular Binary Segmentation (CBS) method and
applied it for CNV detection. \citet{LaiEtal:05} compared 11 algorithms
on the segmentation in array CGH data and reported that CBS is one of
the best. However, they found that CBS was also one of the slowest
algorithms. The computational complexity of these recursive algorithms
are at least $O(n^2)$. \citet{VO:07} proposed a faster CBS algorithm by
setting an early stopping rule. The improved algorithm works much
faster in practice without loss of much accuracy.

Another alternative to exhaustive searching is the $\ell_1$
penalization approach. \citet{HWLZ:05} studied the change-points
detection problem via the following optimization problem:
%e5 ###
%
\begin{equation}\label{l1a}
\mbox{minimize } \Vert \bY-\bmu\Vert ^2\qquad \mbox{subject to } \sum
_j|\mu_j-\mu_{j+1}|\leq s.\vspace*{-2pt}
\end{equation}
\citet{Tibshirani:Wang:2008} applied the fused lasso [\citet
{Tib:fusedlasso:2005}] for change-points detection, which
\[
\mbox{minimizes } \Vert \bY-\bmu\Vert ^2 \qquad\mbox{subject to } \sum_j|\mu
_j|\leq s_1, \qquad\sum_j|\mu_j-\mu_{j+1}|\leq s_2.\vadjust{\goodbreak}
\]
They added one more constraint since they assume that $\bmu$ itself is
a sparse vector.

In fact, the simpler one of the above two, that is,~(\ref{l1a}), is
equivalent to the lasso after a reparametrization $\xi_j=\mu
_{j+1}-\mu
_j$, $j=1,\ldots,n-1$. The complexity of the fastest algorithm to solve
the lasso is at least $O(n^2)$ [\citet{CDA:2007}]. See also recent
developments on the fused lasso [\citet{Rinaldo:2009},
\citet{Hoefling:2010}, \citet{ZhangLange:2010}, \citet{GLasso:2011}].

We consider all aforementioned methods as global methods, in the sense
that they use all data points repeatedly in the process of determining
change-points.
Consequently, the algorithms are usually computationally expensive. To
the best of our knowledge, the algorithms for the change-point
detection that possess consistency properties have the computational
complexities in the order of $O(n^2)$ or higher, although there are
other algorithms that are shown to be fast numerically but whose
theoretical properties in consistency and computational complexities
are not well understood. As pointed out in \citet{BBM00}, the
construction of efficient $O(n)$ algorithms which scale up to large
sequences remains to be an important research topic.

Last but not least, it is worth mentioning that the PennCNV [\citet
{Wangetal:07}] is a fast and efficient algorithm to detect CNVs,
although it is based on a different model from~(\ref{model}).

%s1.3 ###
\subsection{An approach via local information}\label{sec1.3}

To detect the change-points, we need to examine the sequence both
globally and locally. Heuristically, there should exist a neighborhood
around a change-point that contains sufficient information to detect
the change-point. For instance, it is unlikely for the value of
$Y_{10,000}$ to contribute much information for detecting a
change-point around the position $100$. Therefore, global methods
mentioned in the last subsection may not be efficient. By focusing on a
local region, we can reduce the computational burden, especially when
the data set is huge. Based on this philosophy, we propose the
Screening and Ranking algorithm (SaRa), which is a powerful
change-point detection tool with computational complexity down to $O(n)$.

Roughly speaking, the idea is to find a locally defined diagnostic
measure~$D$ to reflect the probability of a position being a
change-point. By calculating the measure $D$ for all positions, which
we call the screening step, and then ranking, we can quickly find a
list of candidates that are most likely to be the change-points. In
fact, such diagnostic measures have been proposed and used in related
literature, for example, \citet{Yin88}, \citet{GPK:99}. For example,
at a point
of interest $x$, a simple diagnostic function is the difference of the
average among observations on the left-hand side and the right-hand side, that
is, $D(x)=\sum_{t=1}^h Y_{j+1-t}/h-\sum_{t=1}^h Y_{j+t}/h$. The larger
$|D(x)|$ is, the more likely that $x$ is a change-point. Note that, to
calculate $D(x)$, we need only the\vadjust{\goodbreak} data points near the position $x$.
Therefore, the strategy through a local statistic can save a lot of
time, compared to global methods.
Moreover, we will show that this fast algorithm is also accurate by
theoretical and numerical studies. In particular, the SaRa satisfies
the sure coverage property, which is slightly stronger than model
consistency.\looseness=1

To summarize, the local methods are usually more efficient than global
ones to solve the change-points problem for high-throughput sequencing
data, and it is feasible to evaluate their computational and asymptotic
properties. With these merits, we believe the local methods have been
under utilized and deserve more attention and development in solving
change-points problems.

This paper is organized as follows. In Section~\ref{sec2} we introduce the
screening and ranking algorithm to solve the change-points problem
(\ref{model}). In Section~\ref{sec3} we study the theoretical properties of the SaRa
and show that it satisfies the sure coverage property. Numerical
studies are illustrated in Section~\ref{sec4}. Proofs of theorems and lemmas are
presented in the supplementary material [\citet{NiuZhangSupp:2012}].

%s2 ###
\section{The screening and ranking algorithm}\label{sec2}

%s2.1 ###
\subsection{\texorpdfstring{Model (\protect\ref{model}) revisited}{Model (1.1) revisited}}\label{sec2.1}

In model~(\ref{model}), assume the mean of each variable in the $j$th
segment is $\beta_j$, that is, $\mu_i = \beta_j,$ for all integers $i
\in(\tau_j, \tau_{j+1}]$.
For clarity, the errors $\varepsilon_i$ are assumed to be i.i.d.
$N(0,\sigma^2)$ as before, although the normality may be relaxed.
Note that in some of the existing work, the step function $\mu$ is
defined as a left-continuous function on unit interval $[0,1]$, and
change-points $0<t_1<\cdots<t_J<1$ are discontinuities of $\mu(x)$. Two
notations are consistent with convention $\mu(\frac in)=\mu_i$ and
$t_j=\tau_j/n$.

For $0 \leq J < \infty$, we refer to the Gaussian model~(\ref{model})
with $J$ change-points as $\mathcal{M}_J$. After a reparameterization
of $\beta_j = \beta_0 + \sum_{l=1}^j \delta_l$, that is, $\delta
_j=\beta
_j-\beta_{j-1}$, $j=1,\ldots,J$, model $\mathcal{M}_J$ has parameters
\[
\bolds{\theta}= (\beta_0, \delta_1, \ldots, \delta_J, \tau_1,
\ldots
, \tau_J, \sigma^2)
\]
to be determined. Let $\bolds{\tau}=(\tau_1,\ldots,\tau_n)$ and
$\bolds{\delta}=(\delta_1,\ldots,\delta_n)$ be the change-points
location vector and jump-size vector, respectively. The purpose is to
estimate $J$ as well as the model parameter vector $\bolds{\theta}$.
Since $n$ is large and $J\ll n$, the variance $\sigma^2$, or an upper
bound of $\sigma^2$, can be easily estimated. Once $J$ and
$\bolds{\tau}$ are well estimated, $\beta_0$ and $\bolds{\delta}$ can be
estimated by the least squares method. In short, the main task is to
estimate the number of change-points $J$ and the location vector
$\bolds{\tau}$.

Roughly speaking, to detect the change-points in model~(\ref{model}),
the SaRa proceeds as follows. In the screening step, we calculate at
each point $x$ a~local diagnostic \mbox{function} $D(x)$ which depends only on
the observations in a small neighborhood $[x-h,x+h]$. In the ranking
step, we rank the local maximum of the \mbox{function} $|D(x)|$ and the top
$\hat{J}$ points $\hat{\tau}_{1}, \ldots, \hat{\tau}_{\hat{J}}$
will be the estimated change-points, where~$\hat{J}$ can be determined
by a thresholding rule or Bayesian information criterion.

In the following subsections, we describe the SaRa in detail.

%s2.2 ###
\subsection{Local diagnostic functions}\label{sec2.2}
We begin with the local diagnostic function which is crucial in the
SaRa. Roughly speaking, an ideal local diagnostic statistic at a
position, say, $x$, is a statistic whose value directly related to the
possibility that~$x$ is a change-point. For model~(\ref{model}), we
first introduce two simple examples of a local diagnostic statistic.

An obvious choice is $D(x)=\sum_{k=1}^h Y_{x+1-k}/h-\sum_{k=1}^h
Y_{x+k}/h$, for a fixed integer $h$ [\citet{Yin88}]. It is simply the
difference between averages of $h$ data points on the left side and
right side of $x$. Heuristically, consider the noiseless case in which
$\bY=\bmu$ is a piecewise constant vector. $D(x)$ is simply a piecewise
linear function, whose local optimizers correspond to the true
change-points. Obviously, the quantity of $D(x)$ reflects the
probability that~$x$ is, or is near to, a change-point.

An alternative choice is the ``local linear estimator of the first
derivative,'' studied in \citet{GPK:99}. Intuitively, if we use a smooth
curve to approximate the true step function $\bmu$, the first
derivative or slope of the approximation curve should be large around a
change-point and near zero elsewhere. The approximation curve as well
as its first derivative can be calculated easily by local polynomial
techniques. Therefore, a local linear estimator of the first derivative
is a reasonable choice for the local diagnostic statistic.

In both examples above, the local diagnostic statistics are calculated
as linear transformations of data points around the position of interest.
We may consider a more general form of local diagnostic functions,
which is the weighted average of $Y_i$'s near the point of interest $x$,
\[
D(x,h) = \sum_{i=1}^n w_i(x) Y_i,
\]
where the weight function $w_i(x)=0$ when $|i-x|>h$.

In the first example, we have
%e6 ###
%
\begin{eqnarray}\label{equalweight}
w_i(x)=\cases{
1/h, & \quad$1-h\leq i-x \leq0,$\vspace*{2pt}\cr
-1/h, & \quad$1\leq i-x \leq h,$\vspace*{2pt}\cr
0, & \quad$\mbox{otherwise.}$}
\end{eqnarray}
To present the weight vector in a kernel, $K(u)$, the weight can be
chosen as $w_i(x)=\operatorname{sgn}(x+\frac12-i)\cdot K_h(i-x)$, where
$K_h(u)=h^{-1}K(u/h)$.
The equal-weight case~(\ref{equalweight}) is a special case
corresponding to the uniform kernel
$K(u)=\frac12\cdot1_{\{|u|\leq1\}}$.\vadjust{\goodbreak}

For the second example, at the point of interest $x$, the local linear
estimator of the first derivative is
\[
D(x,h) = \sum_{i=1}^n w_i Y_i,
\]
where $ w_i = \frac{1}{S_{n,0}S_{n,2} - S_{n,1}^2}K_h(i-x)
[S_{n,0}(i-x)-S_{n,1} ]$,
%$K_h(u) = h^{-1} K(u/h),$
and $S_{n,l} = \sum_{i=1}^n
K_h(i - x)(i-x)^l, l = 0,1,2.$ Note that this is simply the local
linear estimator for the first derivative with kernel $K(u)$ and
bandwidth $h$. See \citet{Fan:local:1996} or \citet{GPK:99}. In the
supplementary material [\citet{NiuZhangSupp:2012}], a class of more
general local diagnostic statistics is introduced.

In all numerical studies within this paper, we use only the
equal-weight diagnostic function~(\ref{equalweight}) for properly
chosen $h$, because of the piecewise constant structure of the mean
$\bmu$. In Section~\ref{mSaRa} we discuss the choices of weight and
bandwidth in detail. From now on, we denote the diagnostic function by
$D(x,h)$ when~$h$ is specified. When the context is clear, we may
simply use $D(x)$.

%s2.3 ###
\subsection{The screening and ranking algorithm}\label{sec2.3}
Given an integer $h$, the local diagnostic function $D(x,h)$ can be
calculated easily. Obviously, a large value of $|D(x,h)|$ implies a
high probability of $x$ being or neighboring a change-point. Therefore,
it is reasonable to find and rank the local maximizers of $|D(x,h)|$.
To be precise, we first define the $h$-local maximizer of a function as follows.

\begin{definition*}
For any number $x$, the interval $(x-h,x+h)$ is called the
$h$-neighborhood of $x$. And, $x$ is an $h$-local maximizer of function
$f(\cdot)$ if $f$ reaches the maximum at $x$ within the
$h$-neighborhood of $x$. In other words,
\[
f(x)\geq f(x')\qquad \mbox{for all } x'\in(x- h,x+ h).
\]

Since we always consider the $h$-local maximizer of $|D(x,h)|$, we may
omit~$h$ when the context is clear. Let $\mathcal{LM}$ be the set of
all local maximizers of the function $| D(x,h) |$. We select a subset
$\hat{\mathcal{J}}=\{ \hat{\tau}_1 < \hat{\tau}_2 < \cdots< \hat
{\tau
}_{\hat{J}} \}\subset\mathcal{LM}$ by a thresholding rule
%e7 ###
%
\begin{equation}\label{thresholding}
|D(\hat{\tau},h)| > \lambda.
\end{equation}

The index set, $\hat{\mathcal{J}},$ and $\hat{J}$ are the SaRa
estimators for the locations and the number of change-points, respectively.
A choice of the threshold $\lambda$ by asymptotic analysis is given in
Section~\ref{sec3}. An alternative way to determine~$\hat{J}$ is by a
Bayesian-type information criterion [\citet{Yao1988}, \citet
{ZhangSiegmund2007}].
By ranking local maximizers of $|D(x)|$, we get a sequence $x_1$,
$x_2,\ldots$ with $|D(x_1)|>|D(x_2)|>\cdots$. Plugging the first
$\tilde
{J}$ $x_i$'s into
\[
\operatorname{BIC}(\tilde{J})= \frac{n}{2}\log\hat{\sigma}^2_{\tilde{J}}+\tilde
{J}\log n,
\]
where $\hat{\sigma}^2_{\tilde{J}}$ is the MLE of the variance assuming
$x_1,\ldots,x_{\tilde{J}}$ are change-points. Then $\hat{J}=\argmin
\operatorname{BIC}(\tilde{J})$ is the estimate for the number of change-points and
$(\hat\tau_1,\ldots,\hat\tau_{\hat{J}})=(x_{(1)},\ldots,x_{(\hat
{J})})$ is
the estimate for the location vector, where $x_{(\cdot)}$ are the order
statistics of $x_1,\ldots,x_{\hat{J}}$. The procedure for the modified
BIC, proposed in \citet{ZhangSiegmund2007}, is the same as the BIC
except using
\[
m\operatorname{BIC}(\tilde{J})= \frac{n}{2}\log\hat{\sigma}^2_{\tilde{J}}+\frac
32\tilde
{J}\log n+\frac12\sum_{i=1}^{\tilde{J}+1}\log\bigl(x_{(i)}/n-x_{(i-1)}/n\bigr),
\]
where $x_{(0)}=0$ and $x_{(\tilde{J}+1)}=n$.
\end{definition*}

%s2.4 ###
\subsection{Computational complexity}\label{sec2.4}
One of the advantages of the SaRa is its computational efficiency. To
calculate the SaRa estimator, there are three steps: calculate local
diagnostic function $D$, find its local maximizers, and rank the local
maximizers. For the equal-weight case, it takes $O(n)$ operations to
calculate $D$, thanks to the recursive formula
\[
D_h(x+1)=D_h(x)+(2Y_{x+1}-Y_{x-h+1}-Y_{x+h+1})/h.
\]
Moreover, $2n$ comparisons are sufficient to find local maximizers of
$D$. Note that there are at most $n/h$ of $h$-local maximizers by
definition, which implies that only $O(\frac{n}{h}\log\frac{n}{h})$
operations are needed to rank these local maximizers. As $h$ is
suggested to be $O(\log n)$ by asymptotic analysis, $O(\frac{n}{h}\log
\frac{n}{h})=O(n)$. In fact, by the thresholding rule, we may not have
to rank all the local maximizers, which can save more time. Altogether,
it takes $O(n)$ operations to calculate the SaRa estimator.

For the general case, the computational complexity is $O(nh)=O(n\log
n)$. Therefore, the SaRa is more efficient than the $O(n^2)$ algorithms.

%s2.5 ###
\subsection{The multi-bandwidth SaRa}\label{mSaRa}
In the SaRa procedure, involved are several parameters such as
bandwidth $h$, weight $w$ and threshold $\lambda$. Although we give an
asymptotic order of these quantities in Section~\ref{sec3}, it is desirable to
develop a data adaptive technique to determine their values.

Similar to other kernel-based methodologies, the choice of the kernel
function, or the weight, is not as crucial as the bandwidth. In this
paper, since we consider model~(\ref{model}) with piecewise constant
mean $\bmu$, the equal-weight~(\ref{equalweight}) performs well. In
some potential applications, when the mean $\bmu$ is piecewise smooth,
the local linear estimator of the first derivative is more appropriate.

Choosing the optimal bandwidth is usually tricky for algorithms
involving bandwidths. To better understand the bandwidth selection
problem, let us look at the local diagnostic statistic $D(x)$ more
carefully. In fact, $D(x)$ is just a statistic used to test the hypothesis
\[
H_0\dvtx  \mbox{there is no change at }x \quad\mbox{vs}\quad H_1\dvtx \mbox{there is a change at }x.
\]
Consider the simplest case when there is no other change-point nearby.
We have
%e8 ###
%
\begin{equation}\label{dist}
D(x,h)\sim N\biggl(\delta,\frac2h\biggr),
\end{equation}
where $\delta$ is the difference between the means on the left and
right of $x$. Obviously, the larger the $h$ is, the more powerful the
test is. Therefore, when applying the SaRa to detect change-points, we
prefer to use long bandwidth for a change-point at which the mean
function jumps only slightly. However, when using a large bandwidth,
there might be other change-points in the interval $(x-h,x+h)$. Then
(\ref{dist}) does not hold anymore. To lower this risk, we prefer small
bandwidth, especially at a point where the jump size is large. In
summary, the bandwidth is of less concern when the change-points are
far away from each other and the mean shifts greatly at each
change-point. Otherwise, we need to choose the bandwidth carefully.
Moreover, the optimal bandwidths for various positions may be
different. It is not an easy task to determine the optimal bandwidth
for each position when we do not have any prior knowledge of change-points.

The choice of the threshold $\lambda$ is easy if we know in advance
that the jump sizes are more than a constant $c$ for all change-points
or we target on only the change-points where the mean shifts
tremendously. Otherwise, the choice can be tricky. If we do not use a
uniform bandwidth for all points, we may have to use different
thresholding rules for different positions.\looseness=1

From the discussion above, we see that it is necessary to develop a
data-adaptive method for parameter selection to enhance the power of
the SaRa. Here we propose the multi-bandwidth SaRa, which can choose
the bandwidth and threshold implicitly and data-adaptively for each
change-point.
The procedure of multi-bandwidth SaRa is as follows. In Step 1, we
select several bandwidths, say, $h_1,\ldots,h_K$, and run the SaRa for
each of these bandwidths. We can use a conservative threshold, say,
$\lambda_k=C\sqrt{2/h_k}\hat\sigma$ with $C=2$ or 3. Note that
$|D(x,h_k)|$, $k=1,\ldots,K$, typically have different local maximizers,
especially when the signal-to-noise ratio is small. We collect these
SaRa estimators which constitute a candidate pool with a moderate size.
In Step 2, we may apply the best subset selection with a BIC-type
criterion [\citet{Yao1988}, \citet{ZhangSiegmund2007}] to the
candidate pool. For
example,~(\ref{JHat}) can be used to estimate the number and locations
of change-points. That is, the candidate pool has a much smaller size
than $n$ and can cover the true change-points in the sense of Theorem~\ref{th1}
in Section~\ref{sec3}. In practice, we recommend a forward stepwise selection or
backward stepwise deletion instead of the best subset selection when
the candidate pool is still big. For instance, to employ the backward
stepwise deletion method, we delete the elements one by one from the
candidate pool. At each step, we delete the one which leads to the
least increase for $\hat\sigma^2$ in~(\ref{JHat}). The procedure stops
if $\frac{n}{2}\log\hat{\sigma}^2_J+J\log n$ stops decreasing, where~$J$ is the size of the remaining candidate pool.

We may consider that the multi-bandwidth SaRa is a combination of local
and global methods. In the first step, the fast local method is
employed to reduce the dimensionality. In the second step, the global
method is applied for accurate model selection. Moreover, the second
step also serves as a~parameter selection tool for the SaRa. For
example, imagine the case that $j=100$ is a change-point, and the SaRa
with bandwidths $h=10$, 15 and~20 suggests 97, 99 and 105 as the
estimates of the change-point, respectively. The second step of the
multi-bandwidth SaRa, that is, the subset selection procedure with the
BIC-type criterion, might suggest~99 as the change-point. In this case,
the bandwidth is selected implicitly as~15. As a result, the
multi-bandwidth SaRa behaves like the SaRa with the optimal bandwidth
at each change-point, which is also verified by the numerical studies.
Note that we can use the strategy of the multi-bandwidth SaRa for a
single bandwidth~$h$, which usually improves the performance. The
reason is that the SaRa arranges the location estimators in the order
of magnitudes of the local diagnostic functions, while the
multi-bandwidth SaRa can rearrange the order using the best subset
selection by making use of global information, and, hence, the
multi-bandwidth SaRa is more reasonable in most cases. In this
procedure, the threshold parameter is selected implicitly for each
position by the subset selection procedure.

The computation cost of Step 2 in the multi-bandwidth SaRa depends on
the size of the candidate pool, which is highly related to the true
number of change-points $J$. With $J\ll n$, the size of the candidate
pool can be well controlled by setting appropriate threshold $\lambda
_k$. Thanks to the sequential structure of observations, the backward
stepwise selection can be employed for subset selection efficiently. In
applications when $n$ is large, Step 2 will not increase the
computation time significantly. %Therefore, the computational
%complexity of the multi-bandwidth SaRa is approximately $K$ times the
%SaRa.

In practice, we still have to choose bandwidths $h_1,\ldots,h_K$ for the
multi-bandwidth SaRa, although their values are not as crucial as the
one in the original SaRa. The choice of these bandwidths should depend
on the applications, in which certain information may be available. If
not, a default setting of $K=3$, $h_1= \log n$, $h_2=2\log n$ and
$h_3=3\log n$ is recommended.\looseness=1

%s3 ###
\section{Sure coverage property}\label{sec3}
The main purpose of this section is to show that the SaRa satisfies the
sure coverage property. In other words, the union of intervals $\hat
{\mathcal{J}} \pm h$ selected by the SaRa includes all change-points
with probability tending to one as $n$ goes to infinity. The
nonasymptotic probability bound is shown as well. It will also be clear
that the sure coverage property implies model consistency of the SaRa.

Let $J=J(n)$ be the number of change-points in model~(\ref{model}).
We denote the set of all change-points by $\mathcal{J}=\{\tau_1<\tau
_2<\cdots<\tau_J\}$. We assume
$\tau_0=0$ and \mbox{$\tau_{J+1}=n$} for notational convenience. For
simplicity, we assume that the noises are i.i.d. Gaussian with variance
$\sigma^2$.
We will use the equally weighted diagnostic function (\ref
{equalweight}) with bandwidth $h$ in this section, but all theorems can
be easily extended for all weights in the family $\mathcal{W}$, defined
in the supplementary material [\citet{NiuZhangSupp:2012}].
We write $\hat{\mathcal{J}}=\hat{\mathcal{J}}_{h,\lambda}$,
defined by
the thresholding rule~(\ref{thresholding}), which depends on the
choices of the bandwidth $h$ and the threshold $\lambda$.

Define $\delta=\min_{1 \leq j \leq J} | \delta_j |$, $L=\min
_{1\leq j\leq J+1}(\tau_j-\tau_{j-1})$. Intuitively, when $\delta
$ and~$L$ are too small, no methods can detect all
change-points. A key quantity for detecting change-points is
$S^2=\delta
^2 L/\sigma^2$. We assume
%e9 ###
%
\begin{equation}\label{assumption}
S^2=\delta^2 L/\sigma^2>32\log n.
\end{equation}

\begin{theorem}\label{th1}
Under Assumption~(\ref{assumption}), there exist $h=h(n)$ and $\lambda
=\lambda(n)$ such that $\hat{\mathcal{J}}=\hat{\mathcal
{J}}_{h,\lambda}
= \{ \hat{\tau}_1, \ldots, \hat{\tau}_{\hat{J}}
\}$ satisfies
%e10 ###
%
\begin{equation}\label{SSP}
\lim_{n \to\infty}\bP(\{\hat{J}=J\}\cap\{ \mathcal{J}
\subset\!\dvtx  \hat{\mathcal{J}} \pm h \}) = 1,
\end{equation}
where by $ \mathcal{J} \subset\!\dvtx  \hat{\mathcal{J}} \pm h $ we mean
$\tau
_j\in(\hat{\tau}_j - h, \hat{\tau}_j + h)$ for all $j\in\{
1,\ldots,J\}$.
In particular, taking $h=L/2$ and $\lambda=\delta/2$, we have
%e11 ###
%
\begin{equation}
\bP(\{\hat{J}=J\}\cap\{ \mathcal{J} \subset\!\dvtx  \hat
{\mathcal{J}} \pm h \}) > 1- 8S^{-1}\exp\{\log n-S^2/32\}.
\end{equation}
\end{theorem}

\begin{remark} We believe Theorem~\ref{th1} can be generalized to the non-Gaussian
case, although we may need a stronger condition than~(\ref{assumption})
in order to bound the tail probability by Bernstein's inequality.
However, the theoretical results are likely to be more complicated.
\end{remark}

\begin{remark} We take $h=L/2$, $\lambda=\delta/2$ and constant $32$ in
Assumption~(\ref{assumption}) for clearer demonstration of our proof.
Obviously the optimal constant in Assumption~(\ref{assumption}) to make
(\ref{SSP}) hold depends on the choice bandwidth $h$, threshold
$\lambda
$ as well as the usually unknown true data generating process.
\end{remark}

\begin{remark} If $L=o(n)$ [e.g., in Assumption~(\ref{assumption}),
$\delta$ and $\sigma$ are constants, $L$ is of order $\log n$], by
Theorem~\ref{th1}, with probability tending to 1, we have
\mbox{$\hat{J}=J$} and $|\hat{\tau}_i-\tau_i|<h$. In particular, $|\hat
{\tau
}_i/n-\tau_i/n|\rightarrow0$, which implies consistency [in the sense
of \citet{Yin88}, \citet{Yao89}] of the SaRa. In fact, the conclusion
of Theorem
\ref{th1} is slightly stronger than model consistency since it shows $|\hat
{\tau
}_i/n-\tau_i/n|\rightarrow0$ uniformly with an explicit rate.
Moreover, in the conventional asymptotic analysis, it is usually
assumed that $J$ is fixed and $\tau_i/n\to t_i$ is constant for each
$i$, as $n$ goes to infinity, which implies $L/n\to$ constant. Our
asymptotic setting~(\ref{assumption}) is more challenging but it
incorporates $L$, $\delta$ and $n$ better.\vadjust{\goodbreak}
\end{remark}

%f1 ###
\begin{figure}

\includegraphics{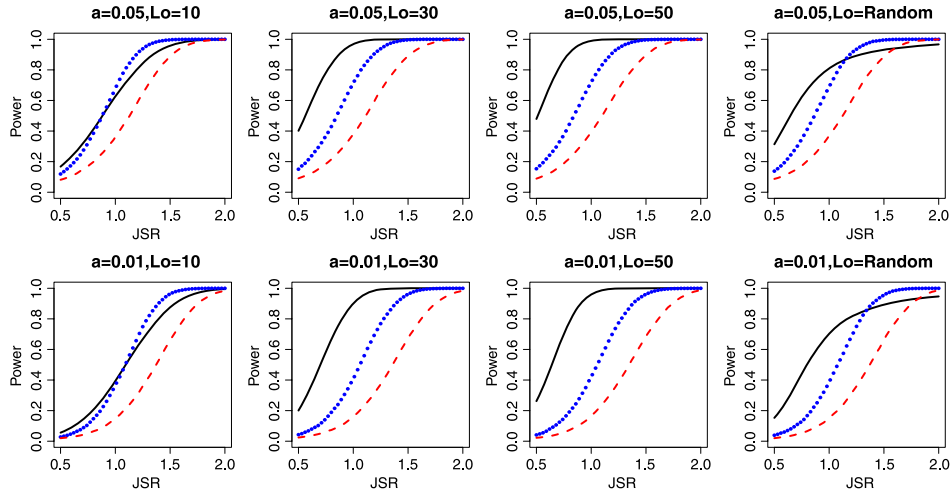}

\caption{Power comparison of the three tests under different Jump size
to Standard deviation Ratios (JSR), change-point locations (Lo) and
type I errors ($\alpha$). Black solid, red dashed and blue dotted lines
correspond to the likelihood ratio test, and SaRa with bandwidths 10
and 15, respectively. The results are based on 10,000
replications.}\label{fig1}\vspace*{-3pt}
\end{figure}

%s4 ###
\section{Numerical results}\label{sec4}

%s4.1 ###
\subsection{Simulation I: Detecting for a single change-point}\label{sec4.1}
Although the SaRa, as a change-point detecting tool, is designed for
large data sets with multiple change-points, we may also compare its
power with classical methods in the conventional setting. In this
subsection, we consider the hypothesis testing problem~(\ref{onetest})
(with a known variance) and compare the SaRa with the likelihood ratio
test~(\ref{SS}).

We assume $Y_i\sim N(\mu_i,\sigma^2)$ with known $\sigma^2$ in
hypothesis testing problem~(\ref{onetest}). Let $n=100$. The maximum of
the diagnostic function\break $\max_{1\leq x\leq
99}|D(x,h)|$ can serve as a test statistic and can be viewed as the
simplest version of SaRa. We consider only the equal weight case (\ref
{equalweight}) for simplicity. We compare three test statistics, the
log likelihood ratio test statistic~(\ref{SS}) and the SaRa with
bandwidths $h=10$ and $h=15$. We run 10,000 replications to get
distributions of the three test statistics under the null hypothesis.
We control the type I error to the levels $\alpha=0.05$ and $0.01,$
respectively, and examine the powers of the three tests when the
change-point locations are 10, 30, 50 and uniformly distributed, respectively.

%t1 ###
%
\begin{table}%[h!]{%\footnotesize
\caption{The estimated model sizes $\hat{J}$ and Sure Coverage
Probabilities (SCP) of SaRa by the thresholding rule under different
settings $(n,L)$ and noise levels $\sigma$. Column 3 lists the
distribution and mean value of the estimated number of change-points.
Columns 4 and 5 list SCPs of two change-points as well as mean distance
between estimated change-point locations and true locations. The
results are based on 1000 replications}\label{Tb:surescreening}
\begin{tabular*}{\textwidth}{@{\extracolsep{\fill}}lccccccc@{}}
\hline
& & \multicolumn{4}{c}{\textbf{Number of change-points}} & & \\[-6pt]
& & \multicolumn{4}{c}{\hrulefill} &  & \\
\multicolumn{1}{@{}l}{$\bolds{(n,L)}$}&
\multicolumn{1}{c}{$\bolds{\sigma}$} & \multicolumn{1}{c}{$\bolds{\hat{J}=2} $} & \multicolumn{1}{c}{$\bolds{<\!2}$} &
\multicolumn{1}{c}{$\bolds{>\!2}$} & \textbf{Mean} & \multicolumn{1}{c}{\multirow{2}{60pt}[11pt]{\centering\textbf{Change-point 1 SCP (Mean)}}} &
\multicolumn{1}{c@{}}{\multirow{2}{60pt}[11pt]{\centering\textbf{Change-point 2 SCP (Mean)}}} \\
\hline
(400, 12) & 0.5\phantom{0} & 63.5\% & 11.7\% & 24.8\%& 2.175 &91.3\% (0.756) &
91.3\% (0.716) \\
& 0.25 & 98.2\% & \phantom{0}1.8\% & \phantom{0}0.0\%& 1.980 &98.9\% (0.129) & 99.1\%
(0.119) \\[3pt]
(3000, 16) & 0.5\phantom{0} & 60.3\% & \phantom{0}8.3\% & 31.4\%& 2.306 &92.8\% (0.814) &
93.4\% (0.776) \\
& 0.25& 98.1\% & \phantom{0}1.9\% & \phantom{0}0.0\% & 1.980 &99.3\% (0.118) & 98.7\%
(0.129) \\[3pt]
(20,000, 20) & 0.5\phantom{0} & 60.2\% & \phantom{0}6.3\% & 33.5\% & 2.343&94.3\% (0.862) &
94.8\% (0.841) \\
& 0.25& 99.3\% & \phantom{0}0.7\% & \phantom{0}0.0\% & 1.993&99.5\% (0.139) & 99.8\% (0.108)
\\[3pt]
(160,000, 24) & 0.5\phantom{0} & 49.5\% & \phantom{0}5.0\% & 45.5\%& 2.599 &95.8\% (0.877) &
95.0\% (1.013) \\
& 0.25& 99.5\% & \phantom{0}0.5\% & \phantom{0}0.0\%& 1.995 &99.8\% (0.096) & 99.7\% (0.148)
\\
\hline
\end{tabular*}
\end{table}

In Figure~\ref{fig1}, we see that the power increases as the Jump size
to Standard deviation Ratio (JSR) increases. The likelihood ratio test
is more powerful when the change-point location is near the middle and
less powerful otherwise. The tests using the SaRa are robust with
respect to the location since the diagnostic function is locally
defined. It is not surprising that the likelihood\vadjust{\goodbreak} ratio test performs
better overall since the SaRa-based tests use information only within a
small neighborhood. However, we see that the SaRa-based tests perform
well when the JSR is greater than 1.5, and the SaRa with $h=15$
outplays the likelihood ratio test when the locations are far from the
middle or random.

%s4.2 ###
\subsection{Simulation II: Sure coverage property}\label{sec4.2}
In this subsection we test the SaRa on a simple but challenging
example. Consider the situation when there is a small CNV with length
$L$ buried in a very large segment with length $n$, where $L=O(\log
n)$. Explicitly, we assume the true mean vector $\bmu=\bE(\mathbf{Y})$
satisfying $\mu_i=0+\delta\cdot I_{\{n/2<i\leq n/2+L\}}$. In other
words, among all~$n$ positions, there are two change-points located on
the position $n/2$ and \mbox{$n/2+L$} with jump sized $\delta$ and $-\delta$.
The locations and the number of aberrations do not influence our
method. We set $(n,L)=(400,12)$, $(3000,16)$, $(20\mbox{,}000,20)$ and
$(160\mbox{,}000,24)$, which satisfy approximately $L \approx2\log n$. We fix
the jump size $\delta=1$ and assume that the noises are i.i.d.
$\mathcal
{N}(0,\sigma^2)$ with $\sigma=0.5$ and $0.25$, so $S^2 \approx8 \log
n$ and $32 \log n$ respectively. We run the SaRa with the thresholding
rule~(\ref{thresholding}), taking $h=\frac34 L$ and $\lambda=\frac34
\delta$. We list the simulation results in Table~\ref
{Tb:surescreening}, based on the 1000 replications. We see that the
SaRa can estimate the number of change-points as well as their
locations accurately in the less noisy case. The two change-points can
be detected with very high probabilities and the estimated errors are
very small with the average below one and median at zero. Even in the
noisy case, the two points can be detected with very high
probabilities, and the number of the false discovered change-points is
very small, at most 1 in most cases. The simulation results match our
theory perfectly.\vadjust{\goodbreak}

%(It would be helpful to add some notes under this table to explain the
%entries. It is not as easy to go back to the text.)

%s4.3 ###
\subsection{\texorpdfstring{Simulation III: An example in Olshen et al. (\citeyear{OVLW:04})}{Simulation III: An example in Olshen et al. (2004)}}\label{sec4.3}
\begin{figure}

\includegraphics{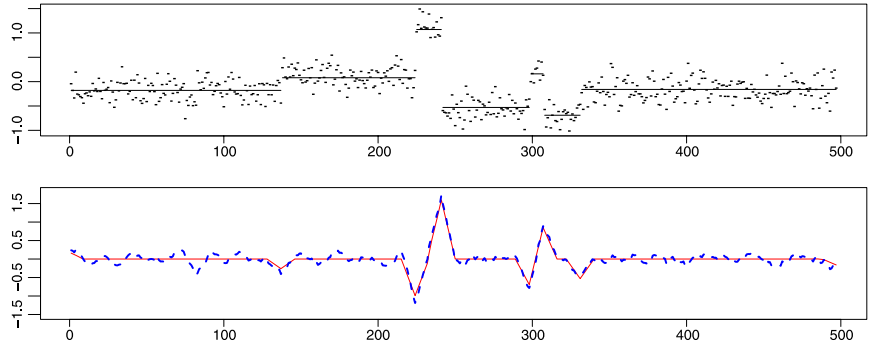}

\caption{Above: example of one Monte Carlo sample of simulation
model (\protect\ref{model:olshen}) with $\sigma=0.2$, $a=0$; bottom: the local
diagnostic functions with bandwidth $h=9$, with noise (blue dashed
line) and without noise (red solid line).}\label{fig2}
\end{figure}

This simulation example, adapted from [\citet{OVLW:04}], is more complex
and realistic. The true data generating process is as follows:
%e12 ###
%
\begin{equation}\label{model:olshen}
Y_i = \mu_i + 0.25\sigma\sin(a\pi i) + \varepsilon_i,\qquad i = 1,
\ldots, n,
\end{equation}
where the error term $\varepsilon$ follows Gaussian distribution
$N(0,\sigma^2)$. The total number of markers $n$ equals 497. There are
six change-points along $\bmu$ with position $\bolds{\tau}=
(137,224,241,298,307,331)$, $\mu_0 =-0.18$, and $\bolds{\delta}=
(0.26,0.99,\break -1.6,0.69,-0.85,0.53)$. A sinusoidal trend was added to
mimic the periodic trend found in the a-CGH data. The noise parameter
$\sigma$ was set to be one of $0.1$ or $0.2$, and the trend parameter
$a \in\{0, 0.01, 0.025\}$ corresponds to none, long and short trends,
respectively.
A simulated data set with $\sigma=0.2$ and $a=0$ is illustrated in
Figure~\ref{fig2}. Among the six change-points, the ones at 137, 298
and 307 are more difficult to detect, since the jump size at 137 is
small, and the length of CNV between 298 and 307 is quite short.

%f2 ###

%t2 ###
%
\begin{table}
\caption{Number of change-points detected by the CBS algorithm and SaRa}\label{Tb:numberOfCP}
%% under different noise levels and trends
%
\begin{tabular*}{\textwidth}{@{\extracolsep{\fill}}lccd{3.0}d{3.0}d{3.0}d{3.0}d{3.0}@{}}
\hline
&&&\multicolumn{5}{c@{}}{\mbox{\textbf{Number of change-points}}}\\[-6pt]
&&&\multicolumn{5}{c@{}}{\hrulefill}\\
\multicolumn{1}{@{}l}{$\bolds{\sigma}$} &\textbf{Trend} & \textbf{Method} & \multicolumn{1}{c}{$\bolds{\leq\!5}$} & \textbf{6} & \textbf{7} & \textbf{8} &
\multicolumn{1}{c@{}}{$\bolds{>\!8}$} \\
\hline
% 0.1 & None & fast CBS & 0 & 921 & 40 & 34 & 5 \\
% 0.1 & None & CBS-SS & 0 & 1000 & 0 & 0 & 0 \\
% 0.1 & None & SaRa (h=9) & 0 & 998 & 2 & 0 & 0 \\
%% 0.1 & None & SaRa (h=12) & 0 & 1000 & 0 & 0 & 0 \\
% 0.1 & None & SaRa (h=15) & 0 & 1000 & 0 & 0 & 0 \\
%% 0.1 & None & SaRa (h=18) & 0 & 1000 & 0 & 0 & 0 \\
% 0.1 & None & SaRa (h=21) & 0 & 1000 & 0 & 0 & 0 \\
% 0.1 & None & m-SaRa & 0 & 1000 & 0 & 0 & 0 \\
% 0.1 & Short & fast CBS & 0 & 776 & 72 & 130 & 22 \\
% 0.1 & Short & CBS-SS & 0 & 1000 & 0 & 0 & 0 \\
% 0.1 & Short & SaRa (h=9) & 0 & 995 & 4 & 1 & 0 \\
%% 0.1 & Short & SaRa (h=12) & 0 & 999 & 1 & 0 & 0 \\
% 0.1 & Short & SaRa (h=15) & 0 & 1000 & 0 & 0 & 0 \\
%% 0.1 & Short & SaRa (h=18) & 0 & 1000 & 0 & 0 & 0 \\
% 0.1 & Short & SaRa (h=21) & 12 & 987 & 1 & 0 & 0 \\
% 0.1 & Short & m-SaRa & 0 & 1000 & 0 & 0 & 0 \\
% 0.1 & Long & fast CBS & 0 & 775 & 114 & 96 & 15 \\
% 0.1 & Long & CBS-SS & 0 & 1000 & 0 & 0 & 0 \\
% 0.1 & Long & SaRa (h=9) & 0 & 997 & 3 & 0 & 0 \\
%% 0.1 & Long & SaRa (h=12) & 0 & 1000 & 0 & 0 & 0 \\
% 0.1 & Long & SaRa (h=15) & 0 & 1000 & 0 & 0 & 0 \\
%% 0.1 & Long & SaRa (h=18) & 1 & 999 & 0 & 0 & 0 \\
% 0.1 & Long & SaRa (h=21) & 6 & 994 & 0 & 0 & 0 \\
% 0.1 & Long & m-SaRa & 0 & 994 & 6 & 0 & 0 \\
0.2 & None & fast CBS & 0 & 872 & 88 & 39 & 1 \\
0.2 & None & CBS-SS & 2 & 998 & 0 & 0 & 0 \\
0.2 & None & SaRa ($h=9$) & 166 & 639 & 150 & 36 & 9 \\
% 0.2 & None & SaRa (h=12) & 84 & 805 & 94 & 15 & 2 \\
0.2 & None & SaRa ($h=15$) & 42 & 901 & 51 & 6 & 0 \\
% 0.2 & None & SaRa (h=18) & 71 & 899 & 28 & 2 & 0 \\
0.2 & None & SaRa ($h=21$) & 157 & 833 & 10 & 0 & 0 \\
0.2 & None & m-SaRa & 0 & 998 & 2 & 0 & 0 \\ \\
0.2 & Short & fast CBS & 0 & 678 & 194 & 101 & 27 \\
0.2 & Short & CBS-SS & 9 & 991 & 0 & 0 & 0 \\
0.2 & Short & SaRa ($h=9$) & 220 & 584 & 156 & 37 & 3 \\
% 0.2 & Short & SaRa (h=12) & 120 & 714 & 143 & 21 & 2 \\
0.2 & Short & SaRa ($h=15$) & 100 & 780 & 107 & 10 & 3 \\
% 0.2 & Short & SaRa (h=18) & 193 & 725 & 73 & 7 & 2 \\
0.2 & Short & SaRa ($h=21$) & 350 & 586 & 60 & 4 & 0 \\
0.2 & Short & m-SaRa & 0 & 992 & 8 & 0 & 0 \\ \\
0.2 & Long & fast CBS & 1 & 695 & 148 & 135 & 21 \\
0.2 & Long & CBS-SS & 6 & 991 & 3 & 0 & 0 \\
0.2 & Long & SaRa ($h=9$) & 263 & 597 & 121 & 15 & 4 \\
% 0.2 & Long & SaRa (h=12) & 141 & 761 & 87 & 11 & 0 \\
0.2 & Long & SaRa ($h=15$) & 101 & 840 & 53 & 6 & 0 \\
% 0.2 & Long & SaRa (h=18) & 181 & 785 & 33 & 1 & 0 \\
0.2 & Long & SaRa ($h=21$) & 317 & 669 & 14 & 0 & 0 \\
0.2 & Long & m-SaRa & 0 & 960 & 40 & 0 & 0 \\
\hline
\end{tabular*}
\end{table}

We applied a few methods to the 1000 simulated data sets of 497 points.
The first method is the fast CBS algorithm by package DNAcopy [\citet
{VO:07}]. Since CBS is quite sensitive and may lead to many false
positives, we also tried the two-stage procedure CBS-SS, which employs
subset selection (SS) with modified BIC [\citet{ZhangSiegmund2007}] to
delete false positives. The second method is the SaRa. We tried a few
different bandwidths ($h=9$, 15 and 21) to test its performance. The
model size was determined by the modified BIC, described at the end of
Section~\ref{sec2.3}. We also applied the multi-bandwidth SaRa (m-SaRa). For the
m-Sara, we first used three bandwidths $h=9$, $15$ and $21$ and chose
the threshold $\lambda=2\sqrt{2/h}\cdot\hat{\sigma}$ for each $h$ to
select candidates.
Then we applied the backward stepwise deletion to get our final
estimate from the set of candidates. (Applying the best subset
selection would be better but a little slower. There are about 20
candidates.) The results for the noisy case ($\sigma=0.2$) are
illustrated in Table~\ref{Tb:numberOfCP}, which lists the number of
change-points detected by these methods under different trends.
We did not assume the noise variance is known, so all the variances in
different scenarios were estimated. To estimate the variance, we first
estimated the mean $\mu_i$ for each point using a local constant
regression. Then the variance can be estimated by the residual sum of
squares divided by $n$.
We omitted the less noisy case $\sigma= 0.1$ since all methods
performed well except the fast CBS, which produced a few false
positives. For $\sigma= 0.2$, the SaRa may underestimate the number of
change-points when the bandwidth is too large or too small. The SaRa
with bandwidth $h=15$ is better than CBS. $h=9$ or $h=21$ is not
optimal since it is hard to detect the first change-point with small
bandwidth and it is difficult to detect the CNV between 298 and 307
with too large bandwidth. However, large bandwidth helps a lot for the
first change-point and small bandwidth gives a more accurate estimate
for the fourth and fifth change-points. It is not surprising that the
multi-bandwidth SaRa, which combines the power of all bandwidths,
performs best in this complex and noisy example. We also observed that
the two-stage procedure CBS-SS can improve the CBS by deleting false
positives. The performances of CBS-SS and m-SaRa are comparable.

In Table~\ref{Tb:DetectRate} we report the detection rate and average
falsely discovered count for the CBS-SS and m-SaRa. We use an
aggressive criterion which considers the change-point undetected if the
estimation error is above 5. A~location estimator is falsely discovered
if there is no true change-point within distance of 5. The detection
rates of two methods are similar. In the cases with trends, the
multi-bandwidth SaRa is slightly better.

%t3 ###
%
\begin{table}
\caption{Detection rates for each change-point and Average Falsely
Discovered (AFD) Count}\label{Tb:DetectRate}
\begin{tabular*}{\textwidth}{@{\extracolsep{\fill}}lcccccccc@{}}
\hline
\textbf{Trend} & \textbf{Method} & \textbf{CP1} & \textbf{CP2} & \textbf{CP3} & \textbf{CP4} & \textbf{CP5} & \textbf{CP6} & \textbf{AFD}\\
\hline
None & CBS-SS & 92.8\% & 100\% & 100\% & 99.8\% & 99.8\% & 99.6\% &
0.076 \\
None & m-SaRa & 90.6\%& 100\% & 100\% & 99.9\% & \phantom{.}100\% & \phantom{.}100\% & 0.097
\\
Short& CBS-SS & 80.9\%& 100\% & 100\% & 99.1\% & 99.1\% & \phantom{.}100\% &
0.191 \\
Short& m-SaRa & 83.0\%& 100\% & 100\% & 99.9\% & \phantom{.}100\% & \phantom{.}100\% & 0.179
\\
Long & CBS-SS & 81.2\%& 100\% & 100\% & 99.7\% & 99.8\% & 97.1\% &
0.217 \\
Long & m-SaRa & 87.1\% & 100\% & 100\% & 99.9\% & \phantom{.}100\% & 99.8\% &
0.172 \\
\hline
\end{tabular*}
\end{table}
%

%s4.4 ###
\subsection{Application to Coriel data}\label{sec4.4}
To test the performance of our algorithm, we apply it to the Coriel
data set which was originally studied by \citet{SnijdersEtal:01}. This
is a typical aCGH data set, which consists of the log-ratios of
normalized intensities from the disease vs control samples, indexed by
the physical location of the probes on the genome. The goal is to
identify segments of concentrated high or low log-ratios. This
well-known data set has been widely used to evaluate CNV detection
algorithms; see \citet{OVLW:04}, \citet{Fridlyand2004132},
\citet{HWLZ:05}, \citet{YinLi:2010},
among others. The Coriel data set consists of 15 fibroblast cell lines.
Each array contains measurements for 2275 BACs spotted in triplicates.
There are 8 whole chromosomal alterations and 15 chromosomes with
partial alterations. All of these aberrations but one (Chromosome 15 on
GM07801) were confirmed by spectral karyotyping.

\begin{figure}

\includegraphics{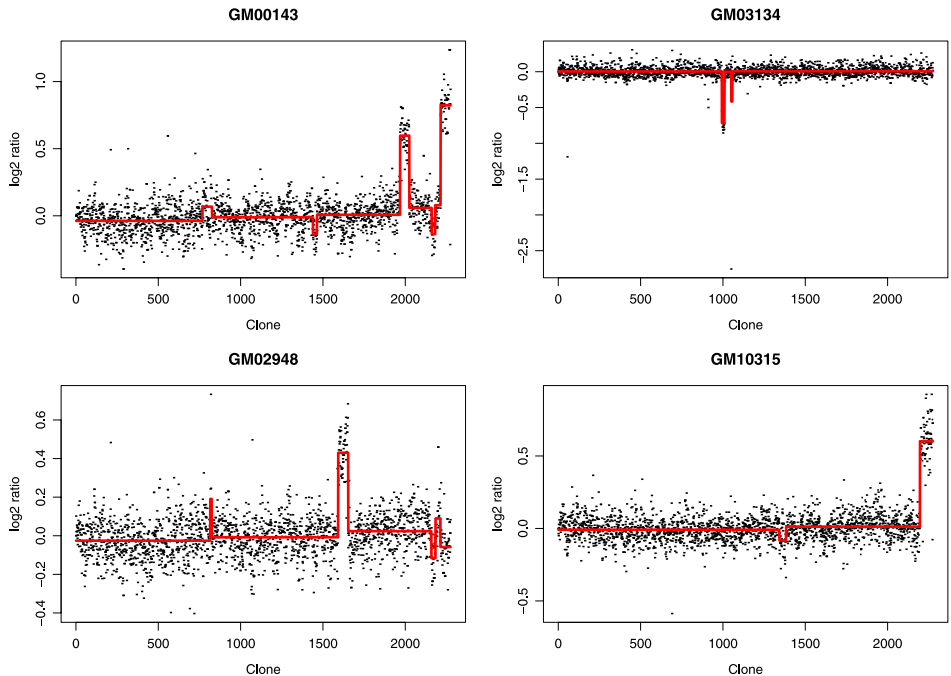}

\caption{Estimated mean functions by the multi-bandwidth SaRa on four
cell lines.}\label{fig4}
\end{figure}

The outcome variable used for the SaRa algorithm was the normalized
average of the $\log_2$ ratio of test over reference. Note that in this
data set there are only three possible mean values corresponding to
loss (monosomy), neutral and gain (trisomy). However, we did not assume
we know this fact in advance. We applied the multi-bandwidth SaRa with
$h=9$, $15$ and $21$ directly to the data set, and used the backward
stepwise deletion with modified BIC to select change-points from all
candidates suggested by the SaRa. As a~result, the multi-bandwidth SaRa
identified all but two alterations (Chromosome 12 on GM01535 and
Chromosome 15 on GM07081). For Chromosome 12 on GM01535, the region of
alteration is represented by only one point and was hence difficult to\vadjust{\goodbreak}
detect. For Chromosome 15 on GM07081, our result agrees with \citet
{SnijdersEtal:01} that there is no evidence of an alteration from the
data set. Our method also found a few alterations that were not
detected by spectral karyotyping. The multi-bandwidth SaRa suggests
that there might be CNVs on chromosomes 7, 11, 21 for GM00143,
chromosome 8 for GM03134, chromosomes 7, 21 for GM02948, and chromosome
11 for GM10315 (see Figure~\ref{fig4}). And most of these additional
CNVs are short aberrations. They may be false positives or true ones
that cannot be confirmed by spectral karyotyping due to its low
resolution. We also tried fast CBS to analyze this data set and found
that CBS selected more than 100 change-points. The result has been
listed in \citet{YinLi:2010}, so we omit it here.

%f3 ###

%s4.5 ###
\subsection{Application to SNP genotyping data}\label{sec4.5}
Due to the low resolution, aCGH techniques are effective to detect only
long CNVs of tens or hundreds of kilobases. By recent studies [\citet
{ConradETAL:2006}, \citet{McCarrollETAL:2006}], it has been shown that short
CNVs are common in the human genome. The genome-wide SNP genotyping
arrays, which are able to assay half a million SNPs, improve the
resolution greatly and offer a more sensitive approach to CNV detection.

We illustrate the SaRa using SNP genotying data for a
father--mother-offspring trio produced by the Illumina 550K
platform.\vadjust{\goodbreak}
The data set can be downloaded along with the \texttt{PennCNV} package.
The data consist of the measurements of a normalized total signal
intensity ratio called the Log R ratio, that is, calculated by $\log
_2(R_{\mathrm{obs}}/R_{\mathrm{exp}})$, where $R_{\mathrm{obs}}$ is the observed total intensity
of the two alleles for a given SNP, and $R_{\mathrm{exp}}$ is the expected
intensity computed from linear interpolation of the observed allelic
ratio with respect to the canonical genotype clusters [\citet{Peiffer:2006}].
For each subject, the Log R ratios along Chromosomes
3, 11 and~20 are included in the data set. There are 37,768, 27,272 and
14,296 SNPs in Chromosomes 3, 11 and 20, respectively. Similar to the
aCGH data, the segments with concentrated high or low Log R ratios
correspond to gains or losses of copy numbers.

We employed the SaRa, multi-bandwidth SaRa, CBS, the fused lasso (FL)
and PennCNV to detect CNVs in this data set. We used R packages \texttt
{DNAcopy} (version 1.14.0), \texttt{cghFLasso} (version 0.2-1) and free
software \texttt{PennCNV} for the last three algorithms. The SaRa and
m-SaRa were implemented by the R program. For the SaRa, we took $h=10$
and $\lambda= 2\sqrt{\log{n}}\cdot\sqrt{2/h}\cdot\hat{\sigma}$.
For the
m-SaRa, we used three bandwidths $h=10$, $20$ and $30$ and the
threshold $\lambda=3\sqrt{2/h}\cdot\hat{\sigma}$ in the first
step. The
backward stepwise selection with modified BIC was employed in the
second step.

%t4 ###
%
\begin{table}
\caption{Number of change-points detected for the offspring by
different methods}\label{Tb:NumOfCh}
\begin{tabular*}{\textwidth}{@{\extracolsep{\fill}}lccccc@{}}
\hline
& \textbf{SaRa} & \textbf{m-SaRa} & \textbf{CBS} & \textbf{FL} & \textbf{PennCNV} \\
\hline
Chromosome 3 & 2 & 19 & 46 & 511 & 2 \\
Chromosome 11 & 4 & \phantom{0}2 & 29 & 345 & 4 \\
Chromosome 20 & 4 & \phantom{0}7 & 16 & 143 & 2 \\
\hline
\end{tabular*}
\end{table}

The result for the offspring is listed in Table~\ref{Tb:NumOfCh}. The
fused lasso and CBS detected too many change-points, most of which are
most likely to be false positives. The performance of the SaRa and
PennCNV are similar. In particular, the SaRa can identify all 4 CNVs
found by PennCNV as well as an additional one on Chromosome 20 from
position 5,851,323 to 5,863,922 kilobase. It seems that the m-SaRa did not
perform as well as the SaRa in this example. The reason is two-fold.
First of all, the signal to noise ratio (i.e., jump size to standard
deviation ratio) is large for this data set. Therefore, the simple SaRa
works very well. Second, in the second step of the m-SaRa, the modified
BIC was employed for model selection. The modified BIC assumes a
uniform prior over the parameter space [\citet{ZhangSiegmund2007}],
which may not be satisfied since the CNVs are very short in this example.

All computations were done on a 3.33~GHz Intel(R) Core(TM)2 Duo PC
running the Windows XP operating system. It took about 16~s for the
\texttt{PennCNV} to show the result for all 3 subjects. So the
computation time for each subject is about 5~s. All other algorithms
were operated in R software. The computation time for each algorithm is
listed in Table~\ref{Tb:Time}. All algorithms are fast and practical.
In particular, the SaRa is one of the fastest. Moreover, we observed
that the computation time for the SaRa increases linearly with the
sample size.

Overall, the SaRa and PennCNV are the best among all algorithms. The
fused lasso and CBS mainly target on the aCGH data and may not be
suitable for the SNP genotyping data. The m-SaRa equipped with a proper
Bayesian-type information criterion is potentially useful. In the SaRa,
we used $h=10$ since it is known that the CNVs are quite short. The
thresholds were chosen by asymptotic analysis of the extremal values of
$D(x)$. In fact, it was quite obvious when the local maximizers were
ranked. For the offspring, there were only 10 local maximizers at which
the values of $|D(x)|$ were larger than $0.57$. The values of $|D(x)|$
at other local maximizers were less than $0.26$.

It is surprising that the SaRa and PennCNV can give similar results
since they are based on different models. The advantage of PennCNV is
that it can utilize more information, that is, the finite states of
copy numbers and the B Allele Frequency (BAF) besides the Log R ratio.
The value of the SaRa is its simplicity and generality. The SaRa can be
implemented easily and is potentially useful in other multiple
change-points problems.

%t5 ###
%
\begin{table}
\caption{Computation time (in seconds) for different methods}\label
{Tb:Time}
\begin{tabular*}{\textwidth}{@{\extracolsep{\fill}}lccccc@{}}
\hline
&\textbf{Number of SNPs} & \textbf{SaRa} & \textbf{m-SaRa} & \textbf{CBS} & \textbf{FL} \\
\hline
Chromosome 3 &37,768 & 1.63 & 8.93 & 63.28 & 3.10 \\
Chromosome 11 &27,272 & 1.16 & 2.64 & 34.05 & 1.79 \\
Chromosome 20 &14,296 & 0.61 & 1.67 & 17.37 & 0.66 \\
\hline
\end{tabular*}
\end{table}

%s5 ###
\section{Discussion}\label{sec5}
Motivated by copy number variation detection, many multiple
change-point detection tools have been invented and developed recently.
However, faster and more efficient tools are needed to deal with the
high dimensionality of modern data sets. Different from other
approaches, we propose a screening and ranking algorithm, which focuses
on the local information. It is an extremely efficient method with
computational complexity down to $O(n)$ and suitable for huge data
sets. Moreover, it is very accurate and satisfies the sure coverage
property. Note that, as far as we know, very few theoretical results
have been developed for multiple change-points detection tools.
Besides the efficiency and accuracy, the SaRa is easily implementable
and extendable. For example, since only local information is involved
in the computation, it is easy to extend it to an on-line version,
which may be useful for financial data. In addition, we may use the
SaRa for the heteroscedastic\vadjust{\goodbreak} Gaussian model when the variance is not
constant. We can estimate the variance using local information and take
it into account when calculating local statistics $D(x,h)$. We should
note, however, that both the implementation and theory for the
heteroscedastic variances will be more complicated and will require
further extensive effort.
In the SaRa procedure, the choices of bandwidth $h$ and threshold
$\lambda$ are important. In applications, optimal choices of these
parameters may depend on the positions and jump-sizes of change-points.
In this case, the multi-bandwidth SaRa, which selects parameter
implicitly and data-adaptively, is practically useful. Knowledge about
the data can be particularly useful in determining reasonable or
meaningful gaps between change-points and and jump-sizes at change
points, which can prevent us from using poor choices of bandwidth~$h$
and threshold $\lambda$.

The SaRa is a useful method for solving change-points problems for a
high dimensional sparse model such as~(\ref{model}) and can be
generalized to solve more general change-point problems. However, we
should point out that the SaRa can be improved for CNV detection since
model~(\ref{model}) may not capture all the characteristics of the CNV
problem. For example, CNVs can be short. Then, it is better to modify
the SaRa accordingly to improve its power.

\begin{supplement}[id=suppA]
\stitle{A description of general weight functions and technical proofs.}
\slink[doi]{10.1214/12-AOAS539SUPP} %[doi,text={...}] - jei reikia
%suskaldyti doi
\slink[url]{http://lib.stat.cmu.edu/aoas/539/supplement.pdf}
\sdatatype{.pdf}
\sdescription{The pdf file contains a description of general weight
functions and the proof of Theorem~\ref{th1}.}
\end{supplement}
%
%%\sname{Supplement A}

% imsref loaded by akundreckaite, 2012-03-13 11:17:55
%

\printaddresses

\end{document}